\documentclass[aip,amsmath,preprint,reprint]{revtex4-1}
\usepackage{hyperref}
\usepackage{graphicx}

\renewcommand{\vec}[1]{{\rm\bf #1}}
\newcommand{\de}{\,\mathrm{d}}
\newcommand{\te}{{\mathsf{T}}}

\begin{document}

\title{Nonlocal van der Waals density functional made faster}

\author{Dimitri N. Laikov}
\email[]{laikov@rad.chem.msu.ru}
\affiliation{Chemistry Department, Moscow State University,
119991 Moscow, Russia}

\date{\today}

\begin{abstract}
A simplification of the VV10 van der Waals density functional [J. Chem. Phys. \textbf{133}, 244103 (2010)]
is made by an approximation of the integrand of the six-dimentional integral
in terms of a few products of three-dimensional density-like distributions and potential-like functions of the interelectronic distance only,
opening the way for its straightforward computation by fast multipole methods.
An even faster computational scheme for molecular systems is implemented where the density-like distributions
are fitted by linear combinations of usual atom-centered basis functions of Gaussian type
and the six-dimensional integral is then computed analytically, at a fraction of the overall cost of a typical calculation.
The simplicity of the new approximation is commensurate with that of the original VV10 functional,
and the same level of accuracy is seen in tests on molecules.
\end{abstract}

\maketitle

\section{Introduction}

Density functional theory is where the worlds of mathematics, physics, and chemistry meet in the most fruitful way,
the rigorous theorems are followed by models
ranging from approximations and interpolations between known limits down to empirical parametrizations of all kinds,
to come up with a computational tool that helps the chemist to understand the structure and behavior of molecular systems
and even to design new ones. Starting from the uniform electron gas~\cite{HK64,KS65},
the generalized-gradient approximations~\cite{LM83,B86,PW86,PBE96} have brought the accuracy to a chemically meaningful level;
a mixing of exact exchange~\cite{B93a,B93b}
or an even better long-range exchange correction~\cite{ITYH01,HSE03,HJS08,L19c}
is the next step in reaching near chemical accuracy.
What is still missing is the small but sometimes important dispersion (also known as van der Waals) energy contribution
for which density functionals evolved~\cite{RLLD00,RDJSHSLL03,DRSLL04}
in the form of a six-dimensional integral ---
a line of further simplifications~\cite{VV08,VV09,VV10}
(notwithstanding formal controversies~\cite{LL10,VV10a})
has led to an elegant analytical expression for its integrand ---
but even then its direct numerical evaluation is rather time-consuming.
In plane-wave methods a fast Fourier transform techniques~\cite{RS09,SGG13} can help,
but here we want to deal with isolated molecular systems
without any further atoms-in-molecules~\cite{GC09,SC10} kind of approximation.

Here we report a new way we have found to simplify the evaluation
of the six-dimensional integral of dispersion energy functionals
by approximating the integrand in terms of a few products of density-like distributions
and potential-like functions --- making it computable by fast multipole methods~\cite{R85}
and hence also by fast density-fitting techniques.
This has been done starting from the VV10~\cite{VV10} functional,
but the same can be applied to other functionals of this kind.
We have implemented its density-fitting version
into our molecular electronic structure code based on traditional Gaussian-type~\cite{B50} functions
and tested the accuracy of the approximation.

\section{Theory}

A family of density functionals for dispersion energy correction have the form of a double space integral
\begin{equation}
\label{eq:e6}
E_6 = \int \int f(\vec{r}_1, \vec{r}_2) \de^3 \vec{r}_1 \de^3 \vec{r}_2
\end{equation}
over a function
\begin{equation}
f(\vec{r}_1, \vec{r}_2) = f\bigl(|\vec{r}_1 - \vec{r}_2|, \rho(\vec{r}_1), \gamma(\vec{r}_1), \rho(\vec{r}_2), \gamma(\vec{r}_2) \bigr)
\end{equation}
of the interelectronic distance $|\vec{r}_1 - \vec{r}_2|$
and the electron density $\rho(\vec{r})$
and its gradient norm $\gamma(\vec{r}) \equiv \bigl|\nabla\rho(\vec{r})\bigr|$ at both points $\vec{r}_1$ and $\vec{r}_2$.
A uniform-electron-gas compensation term
\begin{equation}
\label{eq:e6u}
E_{-6} = -\int \int f\bigl(|\vec{r}_1 - \vec{r}_2|, \rho(\vec{r}_1), 0, \rho(\vec{r}_1), 0 \bigr) \de^3 \vec{r}_1 \de^3 \vec{r}_2
\end{equation}
is also added to the energy.

Though the function $f(r, \rho_1, \gamma_1, \rho_2, \gamma_2)$ may have a simple enough form,
the integral has to be computed by six-dimensional numerical integration.
We want to find a fast approximation $\tilde{E}_6 \approx E_6$ as a sum of a few integrals
\begin{equation}
\label{eq:e6m}
\tilde{E}_6 = \sum_k \int\int q_{1k}(\vec{r}_1) q_{2k}(\vec{r}_2) u_k \bigl(|\vec{r}_1 - \vec{r}_2|\bigr) \de^3 \vec{r}_1 \de^2 \vec{r}_2
\end{equation}
over the products of density-like distributions $q_{jk}(\vec{r}_1)$
and potential-like functions $u_k \bigl(|\vec{r}_1 - \vec{r}_2|\bigr)$
--- opening the way to fast multipole methods or even analytical integral evaluation.

The original VV10~\cite{VV10} model uses a function of five variables
\begin{eqnarray}
& & f_0 (r, \rho_1, \gamma_1, \rho_2, \gamma_2) \\
&=& -\tfrac34 \rho_1 \rho_2 v_0\bigl(r^2, \omega(\rho_1, \gamma_1), \kappa(\rho_1), \omega(\rho_2, \gamma_2), \kappa(\rho_2) \bigr), \nonumber
\end{eqnarray}
and two spatial distributions
\begin{eqnarray}
\label{eq:om}
\omega(\vec{r}) &=& \omega\bigl(\rho(\vec{r}), \gamma(\vec{r}) \bigr), \\
\kappa(\vec{r}) &=& \kappa\bigl(\rho(\vec{r}) \bigr),
\end{eqnarray}
parameterized as
\begin{equation}
\omega(\rho,\gamma) = \left(\frac{4\pi}{3}\rho + C \frac{\gamma^4}{\rho^4}\right)^{1/2},
\end{equation}
\begin{equation}
\kappa(\rho) = B \rho^{1/6},
\end{equation}
$B\equiv\tfrac12 3^{2/3} \pi^{5/6} b$, with two adjustable parameters $C$ and $b$,
and the function of the squared distance $s$ is
\begin{eqnarray}
\label{eq:v0}
& & v_0 (s, \omega_1, \kappa_1, \omega_2, \kappa_2) \\
&=& \frac{1}{(\omega_1 s + \kappa_1)(\omega_2 s + \kappa_2)\bigl((\omega_1 + \omega_2) s + \kappa_1 + \kappa_2 \bigr)} . \nonumber
\end{eqnarray}
The compensation term~(\ref{eq:e6u}) is then simply
\begin{equation}
\label{eq:e6u1}
E_{-6} = -\frac{3^{3/4}}{32 b^{3/2}} \int \rho(\vec{r}) \de^3 \vec{r}.
\end{equation}

Our first approximation assumes $\omega_1 \approx \omega_2$,
helping to simplify the function of Eq.~(\ref{eq:v0}) down to three variables
\begin{eqnarray}
\label{eq:f3}
& & f_1 (r, \rho_1, \gamma_1, \rho_2, \gamma_2) \\
&=& -\tfrac38 \eta(\rho_1, \gamma_1) \eta(\rho_2, \gamma_2) v \bigl(r^2, \mu(\rho_1, \gamma_1), \mu(\rho_2, \gamma_2) \bigr), \nonumber
\end{eqnarray}
with two new spatial distributions
\begin{eqnarray}
\label{eq:ee}
\eta(\vec{r}) &=& \frac{\rho(\vec{r})}{\omega^{3/2}\bigl(\rho(\vec{r}),\gamma(\vec{r})\bigr)}, \\
\label{eq:es}
\mu(\vec{r}) &=& \frac{\kappa\bigl(\rho(\vec{r})\bigr)}{\omega\bigl(\rho(\vec{r}),\gamma(\vec{r})\bigr)},
\end{eqnarray}
and the new function of the squared distance
\begin{equation}
\label{eq:v}
 v (s, \mu_1, \mu_2) = \frac{1}{(s + \mu_1)(s + \mu_2)\bigl(s + \tfrac12 (\mu_1 + \mu_2) \bigr)} .
\end{equation}
Besides greater simplicity, Eq.~(\ref{eq:f3}) allows the interpretation
of $\eta(\vec{r})$ as a density-like quantity
and leads to the geometric mean rule for the $C_6$ coefficients
\begin{equation}
C_6 = \tfrac38 \left(\int \eta(\vec{r}) \de^3 \vec{r} \right)^2,
\end{equation}
whereas the length-scale of the potential-like function $v (s, \mu_1, \mu_2)$
is controlled by the distribution of $\mu(\vec{r})$.

The assumption $\omega_1 \approx \omega_2$ can be justified
by the distribution of $\omega(\vec{r})$ values of Eq.~(\ref{eq:om})
in typical molecules that can be studied numerically,
and in the same way we see
a rather narrow distribution of $\mu(\vec{r})$ values of Eq.~(\ref{eq:es}).

Our next approximation deals with the function~(\ref{eq:v})
\begin{equation}
v(s, \mu, \nu) \approx \tilde{v}(s, \mu, \nu, \beta)
\end{equation}
and is of the form
\begin{equation}
\label{eq:vm}
\tilde{v}(s, \mu, \nu, \beta) = \frac{1}{(s + \beta)^3} + \sum_{m=4}^n \frac{c_m(\mu, \nu, \beta)}{(s + \beta)^m} ,
\end{equation}
where $\beta$ should be a system-independent parameter.
At first we hoped the simplest one with only the first term might be ideal:
it has the $r^{-6}$-tail the dispersion functionals are all about,
and only one spatial distribution~(\ref{eq:ee}) is needed.
This hope was forlorn after the tests on noble-gas homo- and heterodimers:
the value of $\beta$ had to vary from system to system by more than twofold
to reproduce well enough the potential curves.
Thus the length-scale distribution~(\ref{eq:es}) should be somehow accounted for.

From the asymptotic analysis
\begin{eqnarray}
\lim_{s\to\infty} v(s, \mu, \nu) &=& \frac1{s^3} - \frac{3 (\mu + \nu)}{2s^4} + \dots,\\
\lim_{s\to\infty} \tilde{v}(s, \mu, \nu, \beta) &=& \frac1{s^3} + \frac{c_4 (\mu, \nu, \beta) - 3\beta}{s^4} + \dots,
\end{eqnarray}
we get the next simplest term
\begin{equation}
\label{eq:c4}
c_4 (\mu, \nu, \beta) = 3\beta - \tfrac32 (\mu + \nu)
\end{equation}
that may work well because it is the tail that matters.

\begin{figure}
\includegraphics[scale=1.0]{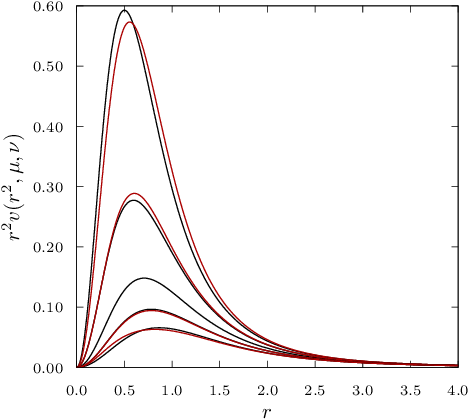}
\caption{\label{fig:v} Function $v(r^2,\mu,\nu)$ and its approximation $\tilde{v}(r^2,\mu,\nu,1)$ of Eq.~(\ref{eq:va})
for $(\mu,\nu) = (\tfrac12,\tfrac12), (\tfrac12,1), (1,1), (1,\tfrac32), (\tfrac32,\tfrac32)$.}
\end{figure}

What also matters is the sum rule
 --- also to leave the term~(\ref{eq:e6u}) unchanged from~(\ref{eq:e6u1}).
From the diagonal ($\mu=\nu $) case,
\begin{eqnarray}
\int\limits_0^\infty r^2 v(r^2,\mu,\mu) \de r &=& \frac{\pi}{16 \mu^{3/2}}, \\
\displaystyle \int\limits_0^\infty r^2 \tilde{v}(r^2, \mu, \mu,\beta) \de r
&=&\displaystyle \frac{\pi}{16\beta^{3/2}} + \frac{\pi c_4 (\mu, \mu, \beta)}{32\beta^{5/2}} \nonumber \\
&+&\displaystyle \frac{5\pi c_5 (\mu, \mu, \beta)}{256\beta^{7/2}},
\end{eqnarray}
and Eq.~(\ref{eq:c4}) we get
\begin{equation}
c_5 (\mu, \mu, \beta) = \frac{16\beta^{7/2}}{5\mu^{3/2}} - 8\beta^2 + \frac{24\beta \mu}{5} .
\end{equation}
For $\mu\ne\nu$ we take the weighted arithmetic and geometric mean
for the last term,
\begin{equation}
\begin{array}{lcl}
c_5 (\mu, \nu, \beta) &=&\displaystyle \frac{16\beta^{7/2}}{5(\mu\nu)^{3/4}} - 8\beta^2 \\
 &+&\displaystyle (1 - \zeta)\frac{24\beta (\mu\nu)^{1/2}}{5} + \zeta\frac{12\beta (\mu + \nu)}{5} ,
\end{array}
\end{equation}
and by equating the derivatives
\begin{eqnarray}
\left. \frac{\partial^2 \tilde{v}}{\partial \mu^2} \right|_{\mu=\nu=\beta} &=&
\frac{3 + \tfrac65 \zeta}{(s + \beta)^5} =
\left. \frac{\partial^2 v}{\partial^2 \mu} \right|_{\mu=\nu=\beta}, \\
\left. \frac{\partial^2 \tilde{v}}{\partial \mu \partial \nu} \right|_{\mu=\nu=\beta} &=&
\frac{3 - \tfrac65 \zeta}{(s + \beta)^5} =
\left. \frac{\partial^2 v}{\partial \mu \partial \nu} \right|_{\mu=\nu=\beta},
\end{eqnarray}
we luckily find
\begin{equation}
\zeta = \frac5{12},
\end{equation}
so our approximation becomes
\begin{equation}
\label{eq:va}
\begin{array}{lcl}
\tilde{v} &=&\displaystyle \frac{1}{(s + \beta)^3} + \frac{\beta}{(s + \beta)^4} \left(3 - \frac{3(\mu + \nu)}{2\beta} \right) \\
&+&\displaystyle \frac{\beta^2}{(s + \beta)^5}
 \left(\frac{16\beta^{3/2}}{5(\mu\nu)^{3/4}} + \frac{14(\mu\nu)^{1/2}}{5\beta} + \frac{\mu + \nu}{\beta} - 8 \right),
\end{array}
\end{equation}
Figure~\ref{fig:v} shows how it works.

Even though $\mu(\vec{r}) \approx\beta$ almost everywhere
(as numerical studies show),
it can be too big or too small somewhere,
for greater numerical stability we prefer to use safer smoothed values
\begin{equation}
\label{eq:mua}
\bar{\mu}(\vec{r}) = \theta\bigl(\mu(\vec{r}),\beta,\alpha \bigr)
\end{equation}
with a filter function
\begin{equation}
\theta(x,c,a) = c + ac \tanh\left(\frac{x - c}{ac} \right)
\end{equation}
(where $c$ sets the typical value and $a$ controls the spread),
with an adjustable parameter $\alpha$ (see below).

(We are well aware that our approximation is (one of) the simplest,
and more accurate ones can be developed applying the idea of Eq.~(\ref{eq:vm})
directly to $v_0$ of Eq.~(\ref{eq:v0}),
\begin{equation}
\label{eq:vn}
\tilde{v}_0 (s, \omega_1, \kappa_1, \omega_2, \kappa_2)
 = \sum_{m=3}^n \frac{c_m(\omega_1, \kappa_1, \omega_2, \kappa_2, \beta)}{(s + \beta)^m} .
\end{equation}
For example, using the approximation
\begin{equation}
\frac2{p + q} \approx \frac2{\sqrt{pq}} - \frac1{2p} - \frac1{2q}
\end{equation}
we might get
\begin{equation}
c_3 = \frac1{\omega_1^{3/2} \omega_2^{3/2}} - \frac1{4 \omega_1^2 \omega_2} - \frac1{4 \omega_1 \omega_2^2},
\end{equation}
and so on, even with $n>5$ in Eq.~(\ref{eq:vn}), but we will not do it here.)

Putting (\ref{eq:va}) into (\ref{eq:f3}) and then (\ref{eq:f3}) into (\ref{eq:e6}) we get a functional of the form~(\ref{eq:e6m}).
With $\eta(\vec{r})$ of Eq.~(\ref{eq:ee}) and $\mu(\vec{r})$ of Eqs.~(\ref{eq:es}) and~(\ref{eq:mua}),
we need four density-like distributions
\begin{eqnarray}
q_0(\vec{r}) &=& \eta(\vec{r}), \\
q_1(\vec{r}) &=& \eta(\vec{r}) \left(\beta^{-1} \mu(\vec{r})\right), \\
q_2(\vec{r}) &=& \eta(\vec{r}) \left(\beta^{-1} \mu(\vec{r})\right)^{1/2}, \\
q_3(\vec{r}) &=& \eta(\vec{r}) \left(\beta^{-1} \mu(\vec{r})\right)^{-3/4},
\end{eqnarray}
and three potential-like functions
\begin{equation}
u_n(r) = \frac{\beta^n}{(r^2 + \beta)^{3 + n}},
\end{equation}
$n=0,1,2$, to compute the approximate dispersion functional
\begin{eqnarray}
\tilde{E}_6 &=& \tilde{E}_6^{(6)} + \tilde{E}_6^{(8)} + \tilde{E}_6^{(10)}, \\
\tilde{E}_6^{(6)} &=& -\tfrac38 U_{00}^{(0)}, \\
\tilde{E}_6^{(8)} &=& -\tfrac38 \left(3 U_{00}^{(1)} - 3 U_{01}^{(1)} \right), \\
\tilde{E}_6^{(10)} &=& -\tfrac38 \left(\tfrac{16}5 U_{33}^{(2)} + \tfrac{14}5 U_{22}^{(2)} + 2 U_{01}^{(2)} - 8 U_{00}^{(2)} \right),
\end{eqnarray}
in terms of the integrals
\begin{equation}
\label{eq:uijn}
U_{ij}^{(n)} = \int\int q_i(\vec{r}_1) u_n \bigl(|\vec{r}_1 - \vec{r}_2|\bigr) q_j(\vec{r}_2) \de^3 \vec{r}_1 \de^3 \vec{r}_2 .
\end{equation}

The numerical evaluation of the six-dimensional integrals~(\ref{eq:uijn})
looks formally like
\begin{equation}
\label{eq:uijnm}
U_{ij}^{(n)} = \vec{q}_i^\te \vec{u}^{(n)} \vec{q}_j .
\end{equation}
One way to do it is by a cubature where the values
\begin{equation}
q_{ki} = w_k q_i(\vec{r}_k),
\end{equation}
\begin{equation}
u_{kl}^{(n)} = u_n\bigl(|\vec{r}_k - \vec{r}_l| \bigr),
\end{equation}
are computed using the points $\{\vec{r}_k \}$
of a three-dimensional integration grid with weights $\{w_k \}$,
the fast multipole methods~\cite{R85} should help here.

Another way is by density fitting
\begin{equation}
q_i(\vec{r}) = \sum_k b_k(\vec{r}) q_{ki}
\end{equation}
where basis functions $\{b_k(\vec{r})\}$ are used
and the coefficients are determined by the least-squares method
\begin{equation}
\vec{q}_i = \vec{S}^{-1} \vec{p}_i
\end{equation}
with the overlap metric
\begin{equation}
S_{kl} = \int b_k(\vec{r}) b_l(\vec{r}) \de^3 \vec{r}
\end{equation}
and the values
\begin{equation}
p_{li} = \int b_l(\vec{r}) q_i(\vec{r}) \de^3 \vec{r}
\end{equation}
computed by numerical integration
\begin{equation}
p_{li} = \sum_k w_k b_l(\vec{r}_k) q_i(\vec{r}_k),
\end{equation}
and we also calculate $q_i(\vec{r}_k)$ from
the fitted~\cite{L97} density $\rho(\vec{r}_k)$ and its gradient $\gamma(\vec{r}_k)$.

The matrix elements in Eq.~(\ref{eq:uijnm}) are
\begin{equation}
u_{kl}^{(n)} = \int b_k(\vec{r}_1) u_n \bigl(|\vec{r}_1 - \vec{r}_2|\bigr) b_l(\vec{r}_2) \de^3 \vec{r}_1 \de^3 \vec{r}_2
\end{equation}
and can be calculated over Gaussian-type~\cite{B50} basis functions
in a way akin to Coulomb integrals~\cite{MD78},
we only have to learn to compute the basic integral
\begin{equation}
\begin{array}{l}
\displaystyle \int\int \frac{\exp\left(-a_1|\vec{r}_1 - \vec{R}_1|^2 -a_2|\vec{r}_2 - \vec{R}_2|^2\right)}
{\bigl(|\vec{r}_1 - \vec{r}_2|^2 + \beta\bigr)^{n + 1}} \de^3 \vec{r}_1 \de^3 \vec{r}_2 \\
= \displaystyle \frac{\pi^3}{n!\, a_1^{3/2} a_2^{3/2} \beta^{n + 1}}
 U_{0n} \left(\frac{|\vec{R}_1 - \vec{R}_2|^2}{\beta}, \frac{a_1 a_2}{a_1 + a_2}\beta\right)
\end{array}
\end{equation}
and its partial derivatives with respect to $\vec{R}_1$ and $\vec{R}_2$,
which can be done in terms of the special functions in two variables
\begin{equation}
U_{mn} (x, y) = (-1)^m \frac{\partial^m}{\partial x^m} U_{0n} (x, y),
\end{equation}
\begin{equation}
\label{eq:umn}
U_{mn} (x, y) = \int\limits_0^\infty \frac{y^{m + 3/2} t^{m + n}}{(y + t)^{m + 3/2}}
 \exp\left(-\frac{xyt}{y + t} - t \right) \de t.
\end{equation}
Our experience with the Boys~\cite{L25} functions
may inspire the work on global approximations to functions~(\ref{eq:umn}),
though for now we evaluate them by one-dimensional numerical integration
using the double-exponential-like quadrature formula
\begin{equation}
\int\limits_0^\infty p(t) \de t = \frac1N \sum\limits_{k=1}^{N - 1} p\bigl(\tau(k/N)\bigr) \tau'(k/N),
\end{equation}
\begin{equation}
\tau(z) = \frac{1}{1 - z} \exp\left(-\frac{a}{z}\right),
\end{equation}
with the parameter
\begin{equation}
a \approx 0.56714329040978387299996866221035555
\end{equation}
being the solution of the equation
\begin{equation}
a = \frac{1 + a}{1 + \exp(a)}.
\end{equation}

\section{Calculations}

Homo- and heteroatomic noble-gas dimers are
our favorite model system for testing dispersion functionals.
We use our long-range-corrected version~\cite{L19c} of the PBE~\cite{PBE96} exchange functional,
our two-component scalar-relativistic approximation~\cite{L19a},
and our \texttt{L2a\_3} basis set~\cite{L19b},
and calculate the values shown in Table~\ref{tab:aa}.

\begingroup
\begin{table*}[t]
\caption{\label{tab:aa}Tests on noble-gas dimers.}
\begin{ruledtabular}
\begin{tabular}{lrrrrrrrr}
system & \multicolumn{2}{c}{VV10} & \multicolumn{2}{c}{Eq.~(\ref{eq:f3})} & \multicolumn{2}{c}{Eq.~(\ref{eq:va})} & \multicolumn{2}{c}{Eqs.~(\ref{eq:va}) and~(\ref{eq:mua})} \\
\cline{2-3}
\cline{4-5}
\cline{6-7}
\cline{8-9}
       & $r$ & $\mathcal{E}$ & $\Delta r$ & $\Delta\mathcal{E}$ & $\Delta r$ & $\Delta\mathcal{E}$ & $\Delta r$ & $\Delta\mathcal{E}$ \\
\hline
HeHe & 6.050 &  22.1 & -0.007 &   1.2 & -0.198 &  1.0 & -0.149 &  1.3 \\
HeNe & 5.911 &  59.4 & -0.012 &   4.8 & -0.067 &  6.9 & -0.036 &  8.0 \\
NeNe & 5.976 & 141.6 & -0.013 &  11.5 & -0.071 & 25.1 & -0.054 & 25.3 \\
HeAr & 6.703 &  78.2 & -0.014 &   6.6 &  0.028 &  2.7 &  0.005 &  5.1 \\
NeAr & 6.683 & 206.7 & -0.023 &  22.0 & -0.045 & 26.4 & -0.042 & 28.6 \\
ArAr & 7.371 & 353.3 & -0.025 &  28.3 & -0.042 & 13.9 & -0.035 & 20.0 \\
HeKr & 7.031 &  82.0 & -0.027 &   9.4 &  0.007 &  5.1 & -0.017 &  8.0 \\
NeKr & 6.952 & 228.5 & -0.033 &  30.9 & -0.062 & 36.9 & -0.059 & 39.5 \\
ArKr & 7.593 & 412.5 & -0.038 &  42.3 & -0.063 & 27.4 & -0.055 & 34.6 \\
KrKr & 7.789 & 494.5 & -0.050 &  61.0 & -0.089 & 47.2 & -0.084 & 55.4 \\
HeXe & 7.522 &  90.4 & -0.067 &  11.7 & -0.089 &  3.8 & -0.093 &  8.0 \\
NeXe & 7.324 & 258.9 & -0.042 &  39.9 & -0.081 & 41.2 & -0.077 & 45.5 \\
ArXe & 7.922 & 493.5 & -0.043 &  55.6 & -0.063 & 31.3 & -0.057 & 41.6 \\
KrXe & 8.099 & 606.9 & -0.056 &  79.4 & -0.098 & 55.4 & -0.088 & 67.4 \\
XeXe & 8.384 & 762.5 & -0.060 & 103.6 & -0.101 & 67.6 & -0.092 & 84.2 \\
\end{tabular}
\begin{flushleft}
Bond lengths $r$ (bohr) and energy well depths $\mathcal{E}$ (microhartree),
and their differences $\Delta r$ and $\Delta\mathcal{E}$ from the VV10 reference values.
\end{flushleft}
\end{ruledtabular}
\end{table*}
\endgroup

Our first approximations of Eq.~(\ref{eq:f3}) works well,
the errors below $1\%$ for bond lengths and below $16\%$ for bond energies are small enough.
For our next working approximation of Eq.~(\ref{eq:va})
we have found the sweet spot $\sqrt{\beta}=4\mbox{ bohr}$
where the errors are not much worse, mostly below $1.2\%$ ($3.3\%$ for HeHe) for bond lengths
and below $18\%$ for bond energies.
The further safety measures of Eq.~(\ref{eq:mua})
with the dimensionless $\alpha = 4$ have a negligible effect.
With all these approximations, we dub this new functional LPBEVVV,
based on the VV10 acronym and the use of \textit{three} potential-like functions.

Noble-gas dimers are the extreme example where the dispersion interaction dominates,
chemically meaningful molecules and intermolecular complexes suffer much less
from all these approximations.

With our density-fitting basis sets~\cite{L20} at \texttt{L1a} and \texttt{L2a} level,
we have already optimized the geometries of many molecular systems
and found negligible differences compared with the use of the original VV10 dispersion functional.

\section{Conclusions}

Our approximations to the original VV10~\cite{VV10} functional
lead to small errors and at the same time allow much faster calculations
when density-fitting is used.

It has already been (silently) applied in our study~\cite{SLNTBF25} of real-world organic reaction mechanisms.

A plane-wave implementation for solid-state systems should be straightforward
using the known methods~\cite{SGG13}.

\section{Data availability}

The data that support the findings of this study are available within the article.

\bibliography{df6}

\clearpage

\end{document}